\documentclass{article}
\usepackage{arxiv}

\usepackage[utf8]{inputenc} 
\usepackage[T1]{fontenc}    
\usepackage{hyperref}
\usepackage{url}            
\usepackage{booktabs}       
\usepackage{amsfonts}       
\usepackage{nicefrac}       
\usepackage{microtype}      
\usepackage{lipsum}		
\usepackage{graphicx}
\usepackage[backend=biber, style=ieee]{biblatex}
\AtEveryBibitem{%
  \clearfield{issn} 

  \ifentrytype{online}{}{
    \clearfield{url}
  }
}
\usepackage{doi}
\usepackage[nolist,nohyperlinks]{acronym}
\usepackage{bm}
\usepackage{bbm}
\usepackage{amsmath}
\usepackage{mathtools}

\DeclareMathOperator*{\argmin}{arg\,min}
\usepackage[capitalise,sort&compress,noabbrev,]{cleveref}
\usepackage[dvipsnames]{xcolor}
\definecolor{0}{RGB}{255, 255, 255}
\definecolor{rwthblue}{RGB}{0,84,159}
\definecolor{rwthbluelight}{RGB}{65,127,183}
\definecolor{rwthbluelighter}{RGB}{142,186,229}
\definecolor{rwthbluelightest}{RGB}{199,221,242}
\definecolor{rwthblack100}{RGB}{0,0,0}
\definecolor{rwthblack75}{RGB}{100,101,103}
\definecolor{rwthblack50}{RGB}{156,158,159}
\definecolor{rwthblack25}{RGB}{207,209,210}
\definecolor{dark-gray}{gray}{0.35}
\usepackage{tikz}
\usetikzlibrary{shapes,arrows.meta,calc,fit,backgrounds,shapes.multipart, 
 positioning,shapes.geometric}
\usepackage{pgfplots,pgfplotstable}
\pgfplotsset{compat=1.18, legend cell align={left}}
\usepackage{paralist}
\newcommand*\samethanks[1][\value{footnote}]{\footnotemark[#1]}

\newtheorem{definition}{Definition}

\newcommand{\ie}{i\/.\/e\/.,\/~}
\newcommand{\eg}{e\/.\/g\/.,\/~}
\newcommand{\cf}{cf\/.\/~}

\title{Measurement Uncertainty: Relating the uncertainties of physical and virtual measurements.}

\author{ \href{https://orcid.org/0000-0002-6342-8157}{\includegraphics[scale=0.06]{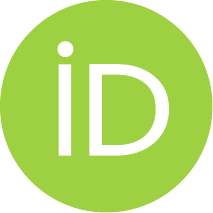}\hspace{1mm}Simon~Cramer}\thanks{Corresponding Author}\,\thanks{Chair of Production Metrology and Quality Management WZL | RWTH Aachen University, Germany}\\
	\texttt{s.cramer@wzl-mq.rwth-aachen.de} \\
	\And
	\href{https://orcid.org/0000-0002-0046-6295}{\includegraphics[scale=0.06]{orcid.pdf}\hspace{1mm}Tobias Müller}
       \samethanks \\
	\texttt{t.mueller@wzl-mq.rwth-aachen.de} \\
	\AND
	\href{https://orcid.org/0000-0002-0011-5962}{\includegraphics[scale=0.06]{orcid.pdf}\hspace{1mm}Robert H. Schmitt}
\samethanks, \thanks{Fraunhofer Institute for Production Technology, Aachen Germany}\\
	\texttt{r.schmitt@wzl-mq.rwth-aachen.de} \\
}

\renewcommand{\shorttitle}{Measurement Uncertainty: Relating the uncertainties of physical and virtual measurements.}

\hypersetup{
pdftitle={Measurement Uncertainty: Relating the uncertainties of physical and virtual measurements.},
pdfauthor={Simon Cramer, Tobias Müller, Robert H. Schmitt},
}

\begin{document}
\maketitle

\begin{abstract}
In the context of industrially mass-manufactured products, quality management is based on physically inspecting a small sample from a large batch and reasoning about the batch's quality conformance.
When complementing physical inspections with predictions from machine learning models, it is crucial that the uncertainty of the prediction is known.
Otherwise, the application of established quality management concepts is not legitimate.	
Deterministic (machine learning) models lack quantification of their predictive uncertainty and are therefore unsuitable.
Probabilistic (machine learning) models provide a predictive uncertainty along with the prediction.
However, a concise relationship is missing between the measurement uncertainty of physical inspections and the predictive uncertainty of probabilistic models in their application in quality management.
Here, we show how the predictive uncertainty of probabilistic (machine learning) models is related to the measurement uncertainty of physical inspections.
 This enables the use of probabilistic models for virtual inspections and integrates them into existing quality management concepts.
 Thus, we can provide a virtual measurement for any quality characteristic based on the process data and achieve a 100\% inspection rate.
 In the field of \emph{Predictive Quality}, the \emph{virtual measurement} is of great interest.
 Based on our results, physical inspections with a low sampling rate can be accompanied by virtual measurements that allow an inspection rate of 100\%.
 We add substantial value, especially to complex process chains, as faulty products/parts are identified promptly and upcoming process steps can be aborted.

\end{abstract}


\begin{acronym}
    \acro{ann}[ANN]{Artificial Neural Network}
    \acro{bnn}[BNN]{Bayesian Neural Network}
    \acro{cad}[CAD]{Computer Aided Design}
    \acro{cdf}[CDF]{Cumulative Distribution Function}
    \acro{doe}[DOE]{Design of Experiments}
    \acro{ffnn}[FFNN]{Feedforward Neural Network}
    \acro{gpr}[GPR]{Gaussian Process Regression}
    \acro{gp}[GP]{Gaussian Process}
    \acro{gum}[GUM]{Guide to the Expression of Uncertainty in Measurement}
    \acro{hmc}[HMC]{Hamiltonian Monte Carlo}
    \acro{hpo}[HPO]{Hyperparameter Optimization}
    \acro{kl}[KL]{Kullback-Leibler}
    \acro{la}[LA]{Laplace Approximation}
    \acro{mae}[MAE]{Mean Absolute Error}
    \acro{map}[MAP]{Maximum a Posteriori}
    \acro{mcmc}[MCMC]{Markov Chain Monte Carlo}
    \acro{mle}[MLE]{Maximum Likelihood Estimation}
    \acro{ml}[ML]{Machine Learning}
    \acro{mpiw}[MPIW]{Mean Prediction Interval Width}
    \acro{nll}[NLL]{Negative Log-Likelihood}
    \acro{nngp}[NNGP]{Neural Network Gaussian Process}
    \acro{pca}[PCA]{Principal Component Analysis}
    \acro{pdf}[PDF]{Probability Density Function}
    \acro{picp}[PICP]{Prediction Interval Coverage Probability}
    \acro{pq}[PQ]{Predictive Quality}
    \acro{pv}[PV]{Process Variable}
    \acro{qc}[QC]{Quality Characteristic}
    \acro{rmse}[RMSE]{Root-Mean-Square Error}
    \acro{rnn}[RNN]{Recurrent Neural Network}
    \acro{svr}[SVR]{Support Vector Regression}
    \acro{vm}[VM]{Virtual Measurement}
    \acro{vi}[VI]{Variational Inference}
    
\end{acronym}
\section{Introduction}
\label{sec:introduction}

In manufacturing, a shift toward more sustainable practices can be observed to cope efficiently with changes in markets and regulation.
A key component in achieving sustainable manufacturing is the assurance of product quality \cite{psarommatisZeroDefectManufacturing2020}.
Therefore, manufacturers continuously monitor the quality of the product, often guided by international standards such as ISO 9001 \cite{DINISO9001}.
The two major quality control strategies are systematic control - every product is inspected - and batch control - when a sample of products from each batch is inspected.
Using inductive statistics, conclusions are deduced from the sample.
Physical inspection of a part is time consuming and expensive.
Therefore, manufacturing companies prefer batch control over systematic control for economic reasons.
However, the major disadvantage of batch control is that defective products can be delivered to the consumer and cause damage to the reputation and costly recalls.
This risk is the $\beta$ error of inductive statistic methods.
Process monitoring and adequate sampling strategies for batch control help reduce this risk.

The basis of every inspection is the physical measurement of a quality characteristic, including tactile, optical, and other methods.
All measurements are inherently uncertain, and this uncertainty propagates to the inspection decision \cite[Chap.~1.1]{muellerModellbildungMittelsSymbolischer2023}.
A measurement result is generally defined as the combination of a measured value and a measurement uncertainty \cite[B.2.11]{Gum}.
This is central for the interpretation of the measurand within the specification limits defined in ISO 14253-1:2017 \cite{DINISO142531}.
Industry guidelines (\eg VDA5) propose procedures for approximating measurement uncertainty \cite[Sec. 3.1]{vdaVDAPrufprozesseignungEignung2011}.
The reference procedures to determine the measurement uncertainty are given in the \ac{gum} \cite{Gum}.
However, because of their extensive nature, they are mostly used in scientific applications.

Due to recent developments in the field of \ac{ml} and virtual metrology, the basis for an inspection can also be a \emph{virtual measurement}.
Interest in virtual measurement has been ever increasing since the first publications in 2005 \cite[Sec.~1]{dreyfusVirtualMetrologyApproach2022}.
The larger goal of virtual metrology is to observe and improve the quality of the process and product using data-driven forecasts.
The basis of most applications is the prediction of quality characteristics, usually called \acl{vm}.
Numerous recent applications can be found in \cite{tercanMachineLearningDeep2022}, and \cite{dreyfusVirtualMetrologyApproach2022}.
Most applications have in common that deterministic (\ac{ml}) models are employed, which lack quantification of their virtual measurement uncertainty.
We found only a few examples where probabilistic models are used, which provide a quantification of their virtual measurement uncertainty, \cf \cite{cramerUncertaintyQuantificationBased2022,yangIndustrialVirtualSensing2020}.
In the larger context, where virtual measurement shall substitute physical measurements, the uncertainty of virtual measurement is essential.
The uncertainty of virtual measurement plays a key role in determining the capability of and building trust in virtual measurements.
With known uncertainty, the risk of wrong decisions can also be quantified in the context of production technology and conformity testing.  
Comparable to a physical measurement, the virtual measurement contributes value once its uncertainty is known.
A virtual measurement plus its uncertainty can be interpreted in established quality management frameworks alongside the (physical) measurements and their measurement uncertainty.
Thus, a systematic quality control of 100\% of all produced parts can be implemented without the expenditure for physical measurements.

 Supplement 1 to the \ac{gum} introduces a three-stage process to determine the measurement uncertainty \cite[Sec.~5.1]{JCGM101}:
\begin{enumerate}
    \item Stage \emph{Formulation}: The measurand $Y \in \mathcal{Y}$ is defined, and the input quantities $\bm{X} \in \mathcal{X}$ on which $Y$ is dependent are determined. Afterwards, the measurement model $f: \mathcal{X} \rightarrow \mathcal{Y}$ is developed, and \acp{pdf} are assigned to the elements $X_i$ of $\bm{X}=(X_1,...,X_N)^T$ based on available knowledge. If any $X_i$ are correlated, a joint \ac{pdf} can be assigned.
    \item Stage \emph{Propagation}: The \acp{pdf} of $\bm{X}$ are propagated through the measurement model $f$ to identify the \ac{pdf} of $Y$.
    \item Stage \emph{Summarizing}: Exploiting the \ac{pdf} of $Y$, one documents the expected value $y:=\mathbb{E}[Y]$, the standard deviation of Y denoted as the standard uncertainty\footnote{$\sqrt[+]{x}$ denotes the positive square root.} $u(Y):=\sqrt[+]{\mathbb{V}[Y]}$, and a coverage interval containing $Y$ with the coverage probability $1-\alpha$.
\end{enumerate}

The overall goal is to associate physical and virtual measurement uncertainty.
To reach this goal we contribute a definition for the virtual measurement and the virtual measurement uncertainty and elaborate on the determination of the virtual measurement uncertainty compared to the process for measurement uncertainty in the \ac{gum}.

In \cref{sec:physical-measurement-uncertainty}, we give some more practical insights on the application of the three stages to determine the measurement uncertainty for physical measurements.
In \cref{sec:virtual-measurement-uncertainty} the three-stage process is applied to virtual measurements based on a probabilistic model to derive the virtual measurement uncertainty in close reference to \cite[Sec.~5]{Gum}.
Finally, we outline the relation between physical and virtual measurement uncertainty in \cref{sec:associate-uncertainty} before concluding in \cref{sec:concluding-remarks}.

\section{Physical Measurement Uncertainty}
\label{sec:physical-measurement-uncertainty}

\begin{figure}
    \centering
    \begin{tikzpicture}[
    label/.style={align=center},
    lowerlabel/.style={label, yshift=-0.45cm},
    upperlabel/.style={label, yshift=0.5cm},
    centerlabel/.style={label, yshift=-0.5cm},
    utriangle/.style={shape=isosceles triangle, fill=rwthblack25, isosceles triangle apex angle=33, minimum height=4cm, rotate=90, inner sep=0cm, outer sep=0cm, anchor=west},
    sbox/.style={rectangle, minimum height=4cm, inner sep=0cm, outer sep=0cm, anchor=south west},
    oosbox/.style={sbox, minimum width=4cm, fill=rwthbluelightest},
    isbox/.style={sbox, minimum width=8cm, fill=rwthbluelighter},
    iuarrow/.style={-{Latex[length=4mm, width'=2.5mm 0.5]}, line width=2mm, rwthblue},
    uarrow/.style={{Latex[length=2mm, width'=2mm 0.5]}-{Latex[length=2mm, width'=2mm 0.5]}, line width=0.75mm}
    ]

    \pgfdeclarelayer{background}
    \pgfdeclarelayer{midground}
    \pgfdeclarelayer{foreground}
    \pgfsetlayers{background,midground,main,foreground}

    \begin{pgfonlayer}{midground}
        \node (triangle1) at (4,0) [utriangle] {};
        \node (triangle2) at (12,0) [utriangle] {};
    \end{pgfonlayer}

    \begin{pgfonlayer}{background}
        \node (box1) at (0,0) [oosbox] {};
        \node (box2) at (4,0) [isbox] {};
        \node (box3) at (12,0) [oosbox] {};
    \end{pgfonlayer}

    \draw[densely dashed] (triangle1.apex)--(triangle1.west);
    \draw[densely dashed] (triangle2.apex)--(triangle2.west);

    \node (oos1) at (box1.north) [upperlabel] {Out of\\Specification};
    \node (lsl) at (box1.north east) [upperlabel] {LSL};
    \node (specification) at (box2.north) [upperlabel] {Specification};
    \node (usl) at (box2.north east) [upperlabel] {USL};
    \node (oos2) at (box3.north) [upperlabel] {Out of\\Specification};

    \node (rmt) at (box2.center) [centerlabel] {Resulting\\Manufacturing\\Tolerance};

    \node (nc1) at (box1.south) [lowerlabel] {Non-\\Conformity};
    \node (uncertainty1) at (box1.south east) [lowerlabel] {Uncertainty};
    \node (conformity) at (box2.south) [lowerlabel] {Conformity};
    \node (uncertainty2) at (box2.south east) [lowerlabel] {Uncertainty};
    \node (nc2) at (box3.south) [lowerlabel] {Non-\\Conformity};

    \draw[uarrow] (triangle1.north)--(triangle1.south) node[pos=0.3, above] {U} node[pos=0.7, above] {U};
    \draw[uarrow] (triangle2.north)--(triangle2.south) node[pos=0.3, above] {U} node[pos=0.7, above] {U};

    \draw[iuarrow] (box3.north east)--(box3.south east) node[pos=0.4, left, align=right, black] {Increasing\\Measurement\\Uncertainty};
    \draw[iuarrow] (box1.north west)--(box1.south west);

    \node (legend) at (box1.south west) [yshift=-1.1cm, align=left, anchor=north west] {\footnotesize LSL / USL: Lower / Upper Specification Limit\\ \footnotesize U: Expanded Measurement Uncertainty};
    \draw (legend.north west)--(legend.north east);

\end{tikzpicture}
    \caption{Limitation of the specification range due to measurement uncertainty according to ISO 14253-1 (see \cite{DINISO142531,muellerModellingComplexMeasurement2020}).}
    \label{fig:uncertainty-tolerance}
\end{figure}
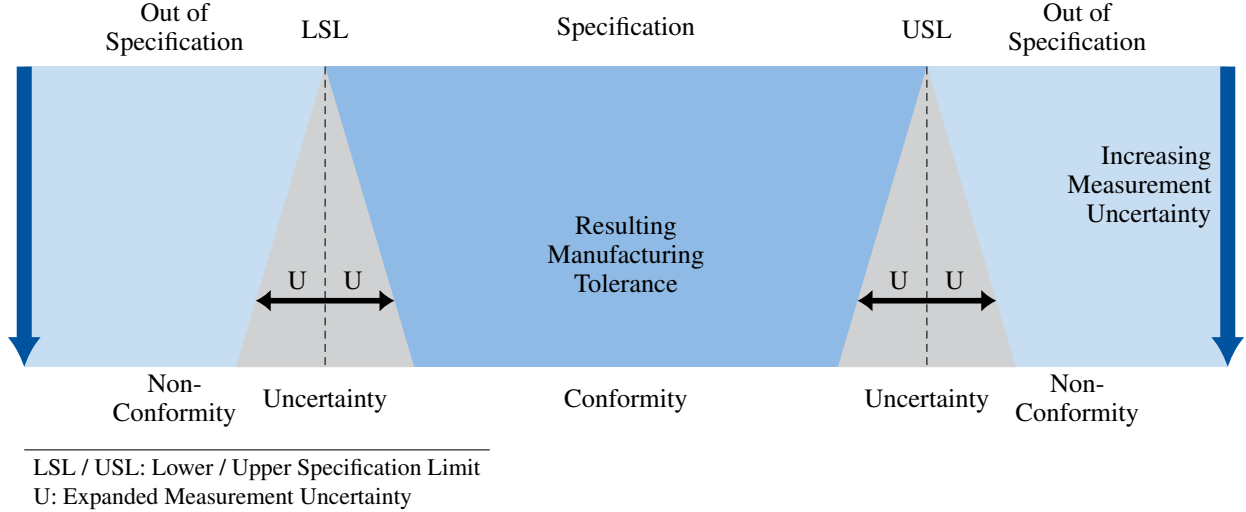

A measurement describes the process of experimentally obtaining one or more quantities (\eg length, energy, electric charge) from a produced part (\cf \cref{def:produced_part}) \cite[Def.~2.1]{bipmInternationalVocabularyMetrology2012}.
A physical measurement is defined here as the use of a physical (measurement) system to obtain the quantity (measurand).
When used in quality management, physical measurements form the basis for a wide variety of decisions.
A classic example of this is the conformity inspection of a produced part $i$ (\eg according to ISO 14253-1).
In conformity inspection, the quality characteristics $Y_i$ of a product are compared to the product specifications (\cf \cref{def:product}).
If the measured values of the quality characteristics are within the specification limits, it is a conformal part; if they are outside, it is a defective part.
Especially at the specification limits, this decision is challenging.
A reason for this is the uncertainty of the measurement $U(Y)$.
Since, by definition, it is impossible to determine the true value of a quantity, every measurement is inherently uncertain \cite[Chap. 1.1]{muellerModellbildungMittelsSymbolischer2023}.
The result of the measurement depends on the measurement procedure, the skill of the operator, the environmental conditions, and various other influences.
For this reason, the conformity decision based on the measurement is also uncertain and can lead to erroneous decisions ($\alpha$- or $\beta$-error).
The higher the measurement uncertainty, or the coarser the measurement uncertainty is estimated, the higher the probability of a wrong decision, and the smaller the remaining range of the specifications in which a reliable decision can be made in favor of a part (see \cref{fig:uncertainty-tolerance}).

To quantify the risk of making an incorrect decision, a determination of the measurement uncertainty is necessary.
The uncertainty of the measurement is a non-negative parameter that characterizes the dispersion of the quantity and describes an interval in which the true value of the measurement lies with a defined probability \cite[Def.~2.26]{bipmInternationalVocabularyMetrology2012}.
The \ac{gum} has been established as the scientific standard for the determination of measurement uncertainty.
It describes a standardized approach to describe the measurement process via the formulation of a mathematical model and, finally, to quantify the measurement uncertainty.
The goal of the \ac{gum} is to establish rules for the evaluation and expression of uncertainty in measurements.
These rules should be applicable for different fields and different uncertainty levels.
The process is divided into three stages, the formulation stage, the propagation stage, and the documentation stage.

\subsection*{Stage 1: Formulation}

In the formulation stage, one defines the measurand $Y$ to be measured and subsequently identifies on which input quantities $\bm{X}=(X_1,...,X_N)^T$ the measurand $Y$ depends.
Typical examples of input quantities are one or more electrical signals from which a geometric property of a produced part is calculated.
In addition, input quantities that characterize the environment can be considered if their influence on the measurement result is known.
Examples of input quantities related to the environment are the temperature of the produced part or the humidity and temperature of the air of the measurement environment.
The input quantities $X_i (1 \leq i \leq N)$ are modeled as random variables whose probability distributions are not necessarily known.

With the desired output and possible inputs prepared, a \emph{measurement model} $f$ in the form of
\begin{equation}
    \label{eq:measurementmodel}
    Y=f(X_1,...,X_N)
\end{equation}
is built that represents the functional relationship between the input quantities and the measurand.
The \emph{measurement model} can be determined analytically (\eg based on physical relationships or information from the literature) or experimentally (\eg based on the execution and evaluation of a \ac{doe}).

The operator assigns respective probability distributions of a common form (\eg Gaussian (normal), rectangular (uniform), etc.) to the input quantities $\bm{X}=(X_1,...,X_N)^T$ based on experience and existing knowledge about the measurement procedure and environmental conditions.
The next step is to determine the probability distribution of $Y$.

\subsection*{Stage 2: Propagation}

In the propagation stage, the probability distributions of the input quantities $\bm{X}$ are propagated through the measurement model $f: \mathcal{X}\rightarrow\mathcal{Y}$ to determine the probability distribution of the measurand $Y$.
In the mathematical sense, the probability distribution of $Y$ is well defined after the formulation phase.
However, to explicitly express \ac{pdf} or \ac{cdf} of the measurand additional steps are necessary.

From \cite[Sec. 5.2]{JCGM101} we know that the \ac{cdf} of $Y$ is
\begin{equation}
    \label{eq:MCMdistributionfunction}
    G_Y(\eta)=\int_{-\infty}^{\eta} g_Y(z)dz,
\end{equation}
where
\begin{equation}
    \label{eq:formaldefinitionPDFofY}
    g_Y(\eta)=\int_{-\infty}^{\infty} \dots \int_{-\infty}^{\infty} g_X(\xi)\delta(\eta-f(\xi)) d\xi_N \dots d\xi_1,
\end{equation}
is a formal definition for the \ac{pdf} of $Y$ with $\delta (\cdot)$ as the Dirac delta function.

We characterize the random variable $Y$ by its expected value $y = \mathbb{E}[Y] = \mathbb{E}[f(X_1,...,X_N)]$ and by its standard uncertainty $u(Y) = \sqrt[+]{\mathbb{V}[Y]}$\footnote{In agreement with \cite[4.1.1 - Note 1]{Gum} the same symbol is used for the physical quantity and the random variable that represents the possible outcomes of an observation of that quantity.}.
According to \cite[Sec. 7]{JCGM104}, there are four different methods to propagate uncertainties through the measurement model.

\paragraph{Analytical Propagation}
The first method is analytical propagation, \ie methods that provide a mathematical representation of the probability distribution for $Y$.
This approach is plausible only in some edge cases.
For example, if the model $f(\cdot)$ is linear and all input quantities are independent, the probability distribution of $Y$ is the convolution of the probability distributions of $X_i (1 \leq i \leq N)$.

However, in general, the calculation of the expectation, the standard deviation, and the coverage interval requires numerical methods, which include a certain degree of approximation \cite[Sec. 5.1.3]{JCGM101}.

\paragraph{First-Order Taylor Approximation}
The second method is the approximation with a first-order Taylor series approximation of $f(\cdot)$.
According to the \emph{law of propagation of uncertainty}, the combined standard uncertainty $u_c(Y)$ is the positive square root of the combined variances $u_c^2(X_i)$ \cite[5.1.2]{Gum}.
\begin{equation}
    \label{eq:gaussianpropagationfirsttaylor}
     u^2(Y) \approx u_c^2(Y)=\sum\limits_{i=0}^{n}\left(\frac{\partial f}{\partial X_i}\right)^2u^2(X_i) 
\end{equation}
The expected value for y is approximated as
\begin{equation}
\label{eq:approxyfortaylorexpansion}
    y = \mathbb{E}[Y] \approx f\left(\mathbb{E}[\bm{X}]\right)
\end{equation}

\paragraph{Higher-Order Taylor Approximation}
For the third method, higher-order terms in the Taylor series expansion are included if the nonlinearity of $f$ is significant.
\begin{equation}
    \label{eq:gaussianpropagationhighertaylor}
    u^2(Y) \approx u_c^2(Y)=\sum\limits_{i=0}^{n}\sum\limits_{j=0}^{n}\left[\frac{1}{2}\left(\frac{\partial^2 f}{\partial X_i\partial X_j}\right)^2+\frac{\partial f}{\partial X_i}\frac{\partial^3 f}{\partial X_i\partial X_j^2}\right]u^2(X_i) u^2(X_j)
\end{equation}
For the approximation of the expected value for $y$ \cref{eq:approxyfortaylorexpansion} is applied.

All conditions of \cite[5.7 and 5.8]{JCGM101} apply to methods two and three, the most prominent being the independence of all $X_i (1 \leq i \leq N)$.
If independence cannot be guaranteed or achieved by reformulation, numerical methods can ease the propagation of uncertainty.

\paragraph{Numerical Methods}
As the fourth approach, the numerical methods are discussed in \cite[Sec. 5.4]{JCGM101}.
The focus is on the Monte Carlo method to propagate the distributions of $X_i$ through the measurement model and approximate $y$ and $u(Y)$. \newline
The Monte Carlo method provides a general method to obtain an approximate numerical representation, denoted $\bm{G}$, of the \ac{cdf} $G_Y(\eta)$ (\cf \cite[Sec. 7.5]{JCGM101}).
The core approach is based on iterative sampling of the \ac{pdf} of $X_i$ and subsequent evaluations $y_r\, (1 \leq r \leq M)$ of the measurement model $f$. \newline
The approximations for $y$ and $u(Y)$ are
\begin{align}
    y      & \approx \frac{1}{M} \sum_{i=1}^{M} y_r     \\
    u^2(Y) & \approx \frac{1}{M-1} \sum_{i=1}^{M} (y-y_r)^2
\end{align}

Since we know the proportionality between the number of samples $M$ and the Monte Carlo error $e_{MC} \sim \frac{1}{\sqrt{M}}$, it is recommended to use large $M$.
The coverage interval for $y$ can be determined by sorting the values of $y_r$, \cf \cite[Sec.~7.7]{JCGM101}.
The output of the Monte Carlo procedure is an estimate for $y$ and an associated standard uncertainty $u(Y)$ based on the coverage interval for $Y$ and the discrete representation of \ac{cdf} through the values of the measurement function.

\subsection*{Stage 3: Documentation}
In the third and final stage, the process of deriving the measurement uncertainty is documented.
This includes the presumed probability distributions of the input quantities, the measurement function, and typically additional knowledge about the boundary conditions of the measurement to repeat and/or verify the process.
Regarding the documentation of the measurement uncertainty, the approximations for the mean value of $y$ and for the standard deviation $u(Y)$, as well as the approximation of \ac{cdf} $G_Y$ are documented.
Concerning the computation of a coverage interval, it is common practice to assume a normal distribution for $Y \sim \mathcal{N}(y,u(Y)^2)$.
Indifferent to the method used for propagation, for a normal distribution, a coverage interval $\left[y-k\,u(Y),y+k\,u(Y)\right]$ with coverage probability $1-\alpha$ can be constructed using the extension factor $k$.
In practice, this interval is often multiplied by a coverage factor $k=2$, so that a coverage probability of approx. $1-\alpha=95\%$ is achieved:
\begin{equation}
    U=k \cdot u(Y)
    \label{eq:coverageinterval}
\end{equation}

The result of the determination of the uncertainty of the measurement is a coverage interval in which the true value of the measurement lies with a defined probability $1-\alpha$.
Based on the uncertainty of the measurement, the specification range can be accordingly limited and the risk of a wrong decision (\eg during conformity inspection) can be quantified.

\subsection*{Summary}
In summary, physical measurements describe the process of obtaining quantities with a physical measurement system.
The measurement uncertainty inherent in measurements is determined (\eg by using the \ac{gum}) in order to quantify the risk of wrong decisions.
Quantification of measurement uncertainty is a standard in industry and science, which creates confidence in the measurement result.
Users can decide on a case-by-case basis whether to take the risk based on the measurement uncertainty.
Although there are approaches that attempt to simplify the determination of measurement uncertainty (including modeling) (VDA 5 or machine learning in modeling) \cite{muellerModellingComplexMeasurement2020,al-abdullahForceTemperatureModelling2018}, the execution of measurements in the production environment still represents a major effort for companies.
The idea of using virtual measurements as an (at least) equivalent alternative to physical measurements helps to reduce the effort of measurements, but the broad application is currently prevented by the fact that existing approaches are commonly deterministic.
There is no information on the reliability of the results, comparable to the uncertainty of measurement.
For this reason, the risk of a wrong decision cannot be quantified based on the results of a virtual measurement.
To enable risk quantification in terms of uncertainty, probabilistic models for virtual measurement and the relation between virtual measurement uncertainty and measurement uncertainty are analyzed in the following.
\section{Virtual Measurement Uncertainty}
\label{sec:virtual-measurement-uncertainty}

Virtual measurements are computed approximations of physical measurements based on data from the manufacturing process, the process data (\cf \cref{def:process-data-bayes}).
Process data can be collected from specifically deployed sensor systems or by recording data from deployed controllers.
The advantages of virtual measurements compared to physical measurements are
\begin{inparaenum}[$(i)$]
    \item that a virtual measurement is easier to acquire and usually available within a few seconds,
    \item that virtual measurement can scale to any amount of produced parts with (almost\footnote{Neglibile, at least in comparison to the cost for physical measurements, \ie expensive measurement devices and trained employees.}) negligible costs for IT resources,
    \item that virtual measurement does not disrupt the flow of parts through the manufacturing process, and 
    \item the virtual measurement is non-destructive to the produced parts.
 \end{inparaenum}

The virtual measurement is usually based on deterministic (\ac{ml}) models.
The models are deterministic in the sense that there is no inherent evaluation of the virtual measurement's uncertainty provided.
However, virtual measurement uncertainty is as important as measurement uncertainty in managing the risk of delivering defective parts to a customer.
To represent the complex dependencies between process data and measurand, \ac{ml} models with thousands to millions of parameters are used \cite{tercanMachineLearningDeep2022}.
Analysis of the uncertainty or sensitivity of these deterministic models is possible in theory \cite[Chap.~10]{sullivanIntroductionUncertaintyQuantification2015}.
Yet, it is unfeasible in practice, as the many degrees of freedom generate extraordinary computational complexity, resulting in long runtimes even on high-performance computing clusters.
Furthermore, as in Stage 1 for physical measurements, strong assumptions regarding the distribution and correlation of the input quantities are required.
An experienced technician/engineer may be able to assume correctly in the controlled environment of a measurement room.
However, if the input quantities are arbitrary process data collected on the shop floor with (sometimes uncalibrated) sensor systems, inferring correct assumptions becomes a daunting task.

The application of probabilistic (\ac{ml}) models to virtual measurements avoids a secondary determination of uncertainty, as these models have an inherent measure of their uncertainty.
Similarly to \cite{JCGM104, Gum, JCGM101}, we focus on modeling a single \ac{qc}.
To model multiple \acp{qc} simultaneously, the model must be extended, depending on whether the correlation between \acp{qc} should be considered \cite[Chap.~1]{jospinHandsonBayesianNeural2021}\cite[Sec.~5.1]{JCGM102}.

We introduce Bayesian interference on a generic model and show how virtual measurements are derived.
The model inputs are \ac{pv} as in \cref{def:process-data-bayes} and the goal is to predict a \ac{qc} as in \cref{def:quality-characterisitic-bayes}.

We postulate a noisy, but known, model between the measurement of a \ac{qc} $Y$ and \acp{pv} $\bm{X}$
\begin{align}
    y(\bm{X}) = f(\bm{X}; \bm{w}) + \epsilon (\bm{X}),
    \label{eq:Bayesian_relationship}
\end{align}
where the mean function $f(\bm{X}; \bm{w}) \in \mathbb{R}$ models the mean \ac{qc} and $\epsilon (\bm{x})  \in \mathbb{R}$ implements the random noise inherent to the production and all involved measurement processes.
The model $f$ is symbolic of the real-world production process and can be queried by producing a new part.
However, $f$ is not known analytically. 

For the noise term, we stipulate a Gaussian probability distribution
\begin{align}
    \epsilon (\bm{X}) \sim \mathcal{N}(0, \sigma_n^2(\bm{X};\bm{w})),
\end{align}
with standard deviation function $\sigma_n(\bm{X};\bm{w}) \in \mathbb{R}^+_0$.

The mean function and standard deviation function are parameterized by the $P \in \mathbb{N}$ model parameters $\bm{w} \in \mathbb{R}^P$.
The relationship in \cref{eq:Bayesian_relationship} implies a Gaussian model density $p(Y|\bm{X}, \bm{w})$ of the \ac{qc} $y$ conditioned on the \acp{pv} $\bm{X}$ and model parameters $\bm{w}$
\begin{align}
    p(Y|\bm{X}, \bm{w}) \sim \mathcal{N}(f(\bm{X}; \bm{w}), \sigma_n^2(\bm{X};\bm{w})),
    \label{eq:single_likelihood}
\end{align}
where
\begin{align}
    \mathbb{E}_{p(Y|\bm{X}, \bm{w})}[Y] & = f(\bm{X}; \bm{w}), \label{eq:exp_val_model_density}    \\
    \mathbb{V}_{p(Y|\bm{X}, \bm{w})}[Y] & = \sigma_n^2(\bm{X};\bm{w}) \label{eq:var_model_density}
\end{align}
are the expectation value and variance, respectively \cite[chapter~2.1]{rasmussenGaussianProcessesMachine2006}.

As parameterization $\bm{w}$ is unknown, we apply Bayesian inference using the knowledge in our database $\mathcal{D}$ (\cf \cref{def:database}).
\begin{equation}
    p(\bm{w}|\mathcal{D}_{\bm{x}}, \mathcal{D}_{y}) = \frac{p(\mathcal{D}_{y}|\mathcal{D}_{\bm{x}}, \bm{w}) p(\bm{w})}{p(\mathcal{D}_{y}|\mathcal{D}_{\bm{x}})}
    \label{eq:bayestheorem}
\end{equation}
Our next step is to combine model density $p(Y|\bm{x}, \bm{w})$ and posterior density $p(\bm{w}|\mathcal{D}_{\bm{x}}, \mathcal{D}_{y})$ to compute predictions.
Consequently, we define the posterior predictive density $p(Y|\bm{X}, \mathcal{D}_{\bm{x}}, \mathcal{D}_{y})$ by averaging the model parameters $\bm{w}$.
\begin{align}
    p(Y|\bm{X}, \mathcal{D}_{\bm{x}}, \mathcal{D}_{y}) & \coloneqq \mathbb{E}_{p(\bm{w}|\mathcal{D}_{\bm{x}}, \mathcal{D}_{y})}[p(Y|\bm{X}, \bm{w})]. \label{eq:posterior_predictive_distribution}
\end{align}
The posterior predictive density specifies the probability distribution of an unobserved \ac{qc} $Y$ for observed \acp{pv} $\bm{X}$ given the knowledge from the provided database $\mathcal{D}$ \cite[chapter~3.3]{Gal2016UncertaintyID}.

The observant reader will notice that this is essentially a measurement result with an associated \ac{pdf} in the sense of \cite[2.9~Note~1]{bipmInternationalVocabularyMetrology2012}.
Therefore, we infer the missing definitions to formally define the \emph{virtual measurement}:

\begin{definition}[Virtual measurement result]
    \label{def:virt-measurement}
    Analog to \cite[2.9~Note~2]{bipmInternationalVocabularyMetrology2012}, a \emph{virtual measurement result} is generally expressed as a single virtual measured quantity value (see \cref{def:virt-measured-quant-value}), and a virtual measurement uncertainty (see \cref{def:virt-measurement-uncertainty}).
\end{definition}

\begin{definition}[Virtual measured quantity value]
    \label{def:virt-measured-quant-value}
    The virtual measured quantity value $\hat{y}(\bm{x}) \in \mathbb{R}$ is the expectation for the unobserved \ac{qc} $Y$ given the \acp{pv} $\bm{X}$. It is defined as the expected value of the posterior predictive density
    \begin{align}
        \hat{y}(\bm{x}) \coloneqq & \mathbb{E}_{p(Y|\bm{x}, \mathcal{D}_{\bm{x}}, \mathcal{D}_{y})}[Y] \label{eq:mean_prediction1}      \\
                         = & \mathbb{E}_{p(\bm{w}|\mathcal{D}_{\bm{x}}, \mathcal{D}_{y})}[f(\bm{x}; \bm{w})] \label{eq:mean_prediction}.
    \end{align}
\end{definition}

In \cref{sec:expectation_value}, a detailed derivation of the equivalence of \cref{eq:mean_prediction1} and \cref{eq:mean_prediction} is given.
Ultimately, the last missing definition is the one of the virtual measurement uncertainty.

\begin{definition}[Virtual measurement uncertainty]
    \label{def:virt-measurement-uncertainty}
    The virtual measurement uncertainty $\hat{\sigma}(\bm{X})$ is defined as the standard deviation of the posterior predictive density
    \begin{align}
        \hat{\sigma}^2(\bm{X})  \coloneqq & \mathbb{V}_{p(Y|\bm{X}, \mathcal{D}_{\bm{x}}, \mathcal{D}_{y})}[Y] \label{eq:predictive_uncertainty1} \\
                                 = & \mathbb{E}_{p(\bm{w}|\mathcal{D}_{\bm{x}}, \mathcal{D}_{y})}[\sigma_n^2(\bm{x}; \bm{w})] + \mathbb{V}_{p(\bm{w}|\mathcal{D}_{\bm{x}}, \mathcal{D}_{y})}[f(\bm{x}; \bm{w})]. \label{eq:predictive_uncertainty}
    \end{align}
\end{definition}
The equivalence of \cref{eq:predictive_uncertainty1} and \cref{eq:predictive_uncertainty} is shown in \cref{sec:variance}.

In application, the process model $f(\cdot, \cdot)$ and the noise term $\epsilon(\cdot)$ are not analytically known.
The available knowledge is given in the database $\mathcal{D}$.
Hence, we adjust our problem formulation to determine the parameters $\bm{w} \in \bm{W}$ of the approximation $H_{\bm{w}}$
\begin{equation}
    Y = f(\bm{X}; \bm{w}) + \epsilon (\bm{X}) \approx H_{\bm{w}}(\bm{x}) = p(Y|\bm{x}, \mathcal{D}_{\bm{x}}, \mathcal{D}_{y}) \quad \forall (x, y) \in \mathcal{D}.
    \label{eq:virtual-measurement-model}
\end{equation}

Once we determine a suitable set of parameters $\bar{w}$, we can predict \ac{qc} from \ac{pv} with the mapping $H_{\bar{w}}$.

We cannot solve the obscured integrals behind \cref{eq:mean_prediction} and \cref{eq:predictive_uncertainty}, because the evidence density $p\left(\mathcal{D}_y |\mathcal{D}_x\right)$ is intractable \cite[Chap.~2.2]{depewegModelingEpistemicAleatoric2019}.
However, it is necessary to normalize the posterior density to ensure $ \int p\left(\bm{w} | \mathcal{D}_y , \mathcal{D}_x\right) d\bm{w} = 1$ (\cf \cref{eq:bayestheorem}).
The problem of intractable evidence density is overcome by the use of appropriate mathematical methods.
After all, we can now implement the three stages as introduced in Supplement 1 to the \ac{gum}.

\subsection*{Stage 1: Formulation}
In the formulation stage, the measurand $Y$ to be virtually measured is defined and subsequently identified on which \ac{pv} the measurand depends.
If available, input quantities characterizing the environment (\eg air temperature) or providing additional information about the machinery (\eg age of the tools) are considered.
\ac{pv} are recorded by a set of sensors \cf \cref{def:process-data-bayes}.
As a basis for virtual measurement, the database $\mathcal{D}$ (\cf \cref{def:database}) is now populated with samples of \ac{pv} and \ac{qc} assigned to a produced part $i$.
As in all applications of machine learning, the quality of the database is a necessary condition for high-performing models.
The probability distributions of the input quantities $\mathbf{X}$ are implicitly given in the database $\mathcal{D}$.

With the desired output and inputs prepared, a \emph{virtual measurement model} $H_{\bm{w}}(x)$ is designed in the form of \cref{eq:virtual-measurement-model}.
In the case of virtual measurement, the operator does not explicitly know or formulate the model $f: \mathcal{X}\rightarrow\mathcal{Y}$.
Instead, an algorithm is selected to instantiate the probabilistic model and to approximate the intractable evidence density with appropriate methods.
Two popular choices are the \ac{mcmc} sampling (\eg \cite[Chap.~5]{betancourtConceptualIntroductionHamiltonian2018}) and Bayesian \ac{vi} (\eg \cite{kucukelbirAutomaticDifferentiationVariational2016}).
All algorithms depend on a set of hyperparameters, which the operator has to choose before the propagation stage.
As hyperparameters have a great influence on the performance of the virtual measurement model, the operator repeats the propagation stage until a suitable set is discovered.

\subsection*{Stage 2: Propagation}

In the propagation stage, the parameters $\bm{w}$ for the posterior predictive density $H_{\bm{w}}(x) = p(X|\bm{x}, \mathcal{D}_{\bm{x}}, \mathcal{D}_{y})$ are calculated.
In the machine learning community, this process is called \emph{model training}.
We introduce Bayesian \ac{vi} as an exemplary method for the propagation of uncertainties.

\ac{vi} additionally approximates the exact posterior density $p(\bm{w}|\mathcal{D}_{\bm{x}}, \mathcal{D}_y)$ with a variational density $q(\bm{w}; \bm{\phi})$. The variational density is parameterized by a vector of variational variables $\bm{\phi} \in \Phi$.
The normalization of the variational density is known so that \ac{vi} does not require knowledge of the intractable evidence density $p(\mathcal{D}_{y}|\mathcal{D}_{\bm{x}})$.
For the variational density, we choose a probability density that is supported by non-Markovian sampling procedures.
For example, samples of a Gaussian variational density can be obtained with a Box-Muller transform \cite[Chap.~5]{jospinHandsonBayesianNeural2021}.
The crucial part of \ac{vi} is the optimization of the variational variables $\bm{\phi}$.
Therefore, our objective is to make the variational density as similar to the posterior density as possible.
We propose the \ac{kl} divergence as a non-negative measure of the dissimilarity between the variational and posterior density \cite[Chap.~3.13]{goodfellowDeepLearning2016}.
\begin{align}
    \mathrm{KL}[q(\bm{w}; \bm{\phi})||p(\bm{w}|\mathcal{D}_{\bm{x}}, \mathcal{D}_{y})] \coloneqq \mathbb{E}_{q(\bm{w}; \bm{\phi})}\left[ \log \frac{q(\bm{w}; \bm{\phi})}{p(\bm{w}|\mathcal{D}_{\bm{x}}, \mathcal{D}_{y})} \right]. \label{eq:def_kl}
\end{align}
The optimal variational variables $\bm{\phi}^*$ are given by the minimum of the \ac{kl} divergence.
The optimization problem for the variational variables is
\begin{align}
    \bm{\phi}^* &= \argmin_{\bm{\phi} \in \Phi} \mathrm{KL}[ q(\bm{w}; \bm{\phi})||p(\bm{w}|\mathcal{D}_{\bm{x}}, \mathcal{D}_{y})] \label{eq:kl_1} \\
    &= \argmin_{\bm{\phi} \in \Phi} \mathcal{F}(\bm{\phi}), \label{eq:kl_2}
\end{align}
where $\mathcal{F}(\bm{\phi})$ denotes the variational free energy
\begin{align}
    \mathcal{F}(\bm{\phi}) \coloneqq \mathrm{KL}[ q(\bm{w}; \bm{\phi})||p(\bm{w})] - \mathbb{E}_{q(\bm{w}; \bm{\phi})}[\log p(\mathcal{D}_y|\mathcal{D}_{\bm{x}}, \bm{w})].
\end{align}
The variational free energy is independent of the intractable evidence density $p(\mathcal{D}_{y}|\mathcal{D}_{\bm{x}})$ \cite[Chap.~3]{blundellWeightUncertaintyNeural2015}.
For a detailed derivation of the equivalence of \cref{eq:kl_1} and \cref{eq:kl_2}, see \cref{sec:kl_divergence}.

\begin{figure}[tb]
    \centering
    \begin{tikzpicture}
    \begin{axis}[
    xmin=0,
    xmax=1,
    ymin=0,
    ymax=1,
    xlabel=Model parameter $w_i$,
    ylabel=Model parameter $w_j$,
    axis equal image,
    legend pos=outer north east,
    ticks=none
    ]
        \addplot [rwthblue, smooth] table {
        x y
        0.18 0.18
        0.18 0.54
        0.4 0.7
        0.8 0.9
        0.8 0.4
        0.7 0.39
        0.6 0.2
        0.23 0.16
        0.18 0.18
        };
        \addlegendentry{Exact posterior density}
        \draw[red, dashed](axis cs:0.5,0.5) ellipse[x radius=0.45, y radius=0.22, rotate=45];
        \addlegendimage{red, dashed}
        \addlegendentry{Correlated approximation}
        \draw[dotted, thick](axis cs:0.5,0.5) circle[radius=0.25];
        \addlegendimage{dotted, thick}
        \addlegendentry{Uncorrelated approximation}
    \end{axis}
\end{tikzpicture}
    \caption[Equiconfidence lines of marginal posterior density and its approximations.]{Equiconfidence lines of marginal posterior density and its approximations. An approximation that considers correlation between the model parameters $w_i$ and $w_j$ can approximate the marginal posterior density more accurately \cite[Chap.~3.1]{kucukelbirAutomaticDifferentiationVariational2016}.}
    \label{fig:surrogate_posterior_correlation}
\end{figure}
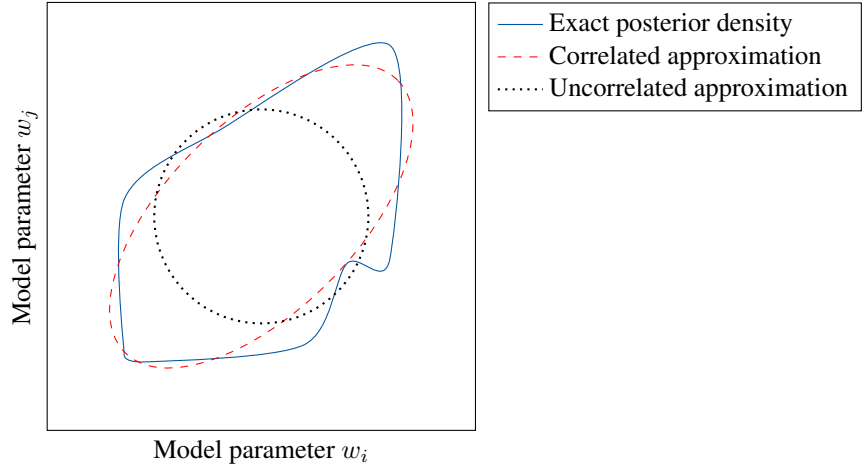

Typically, one limits the variational density to a certain function class, for instance, a Gaussian approximation of the posterior density.
The variational density is given by
\begin{align}
    q(\bm{w}; \bm{\mu}_{VI}, \bm{\Sigma}_{VI}) \sim \mathcal{N} (\bm{\mu}_{VI}, \bm{\Sigma}_{VI})
\end{align}
with mean vector $\bm{\mu}_{VI} \in \mathbb{R}^P$, covariance matrix $\bm{\Sigma}_{VI} \in \mathbb{R}^{P \times P}$, and variational variables as $\bm{\phi} = (\bm{\mu}_{VI}, \bm{\Sigma}_{VI})$.
The covariance matrix of the Gaussian approximation can be restricted to certain forms to decrease the computational requirements. 
A full-rank covariance matrix considers the correlations between all the model parameters $\bm{w}$.
A diagonal covariance matrix disregards correlations and implements a mean-field approach.
Covariance matrices that consider correlations have the potential to approximate the posterior density with greater accuracy.
This phenomenon is illustrated for the marginal posterior densities of two model parameters in \cref{fig:surrogate_posterior_correlation} \cite[Chap.~2.4]{kucukelbirAutomaticDifferentiationVariational2016}.

To compute the expected values and variances in \cref{eq:mean_prediction,eq:predictive_uncertainty} numerically, the model is sampled multiple times.
One can construct a \ac{cdf} of the measurand that is compatible with \cref{eq:MCMdistributionfunction} based on the application of \cite[Sec.~7.4,7.5]{JCGM101}.
For each virtual measurement and independent virtual measurement, uncertainty is computed.

\subsection*{Stage 3: Documentation}
In the third stage, the process of deriving the virtual measurement and its uncertainty is documented.

This includes the database $\mathcal{D}$, or at least descriptive statistical parameters, to monitor new sets of input quantities for concept drift and make necessary model adjustments \cite[Sec.~1]{bayramConceptDriftModel2022}.
Changes on the shop floor, such as the servicing of a workstation or replacement of a tool, can invalidate a virtual measurement model and require retraining.
Furthermore, all hyperparameters regarding the virtual measurement model and the model itself have to be preserved.

To generate virtual measurements for a new produced part, the virtual measured quantity value and the virtual measurement uncertainty must be computed according to \cref{eq:mean_prediction,eq:predictive_uncertainty}.
The sampling-based approximation of \ac{cdf} $G$ can be documented if required.
Concerning the derivation of a coverage interval, it is common practice to assume a normal distribution for $Y \sim \mathcal{N}(\hat{y},\hat{\sigma}^2)$.
Indifferent to the method used for propagation, for a normal distribution, a coverage interval $\left[\hat{y}-k\,\hat{\sigma}^2,\hat{y}+k\,\hat{\sigma}^2\right]$ with coverage probability $1-\alpha$ can be constructed by using the extension factor $k$.
In practice, this interval is often multiplied by a coverage factor $k=2$, so a coverage probability of approximately $1-\alpha=95\%$ is achieved, \cf \cref{eq:coverageinterval}.

The result of the virtual measurement uncertainty is a coverage interval in which the true value of the measurement lies with a defined probability $1-\alpha$ for each produced and virtually measured part.

\subsection*{Summary} 
In summary, virtual measurement describes the process of obtaining quantities with a virtual measurement application.
Using probabilistic models, the virtual measurement's uncertainty is individually determined for each virtual measurement.
Quantification of the uncertainty of virtual measurements is not yet standard in industry or science, so the results induce risk in all upstream decisions.
In the semiconductor industry, where virtual measurement has a greater market penetration than in manufacturing, a sampling decision system is often used to infer how the produced part is processed further \cite[Sect.~1]{kurzMonitoringVirtualMetrology2013}.
\section{Relation of Physical and Virtual Measurement Uncertainty}
\label{sec:associate-uncertainty}

We introduced the physical measurement uncertainty according to the \ac{gum} in \cref{sec:physical-measurement-uncertainty} and the virtual measurement uncertainty in \cref{sec:virtual-measurement-uncertainty}.
Furthermore, to enable accurate discussions and avoid misinterpretation, we introduce definitions of a virtual measured quantity value \cref{def:virt-measured-quant-value}, a virtual measurement uncertainty \cref{def:virt-measurement-uncertainty}, and a virtual measurement \cref{def:virt-measurement}, analogous to the VIM \cite{bipmInternationalVocabularyMetrology2012}.

In the following, we reexamine the three stages to determine the measurement uncertainty.
Indicating similarities, parallels, and deviations, we show the relation of the two types of uncertainty and argue their equivalence for the application in quality management frameworks.

\subsection*{Stage 1: Formulation}

Following the \ac{gum} the main outcomes of the formulation stage are 
\begin{inparaenum}[$(i)$]
   \item a defined measurand $Y \in \mathcal{Y}$, 
   \item the input quantities $\bm{X} \in \mathcal{X}$,
   \item the assignment of \acp{pdf} to the elements $X_i$ of $\bm{X}=(X_1,...,X_N)^T$,
   \item and a designed measurement model $f:\mathcal{X}\rightarrow\mathcal{Y}$.
\end{inparaenum}

The procedure for the virtual and the physical measurements agree on the definition of a measurand $Y \in \mathcal{Y}$.

Regarding the input quantities $\bm{X}$, there are multiple differences between the two procedures. 
For physical measurement, the input quantities come from qualified sources (\ie sensors with known uncertainty), and their selection is dependent on the knowledge of the operator about the measurement process.
Their number is limited by the sensors available to characterize the environment and other effects.
The input quantities for the virtual measurement are recorded from qualified and unqualified sources (\ie sensors with unquantified uncertainty), and their selection is dependent on the knowledge of the operator about the manufacturing process.
The amount of input quantities is restricted by the accessibility of data from the production process.
Often, input quantities from the manufacturing process are time series with high frequencies, which require specialized preprocessing to extract profitable information.
For both measurement types, the selection of input quantities is not exclusive, resulting in a contribution to the overall measurement uncertainty.

With respect to the assignment of \acp{pdf} to the elements $X_i$ of $\bm{X}=(X_1,...,X_N)^T$, virtual and physical measurements follow different philosophies.
For the physical measurement, the operator assigns \acp{pdf} to the input quantities based on experience and knowledge about the measurement procedure and the environment, or based on experiments.
This approach is not feasible for the virtual measurement, as the measurement uncertainty for the sensors on the workstations is rarely determined, changes in the environment occur regularly, and the correlation of input quantities depends on the state of the manufacturing process.
Hence, the distribution and correlation of input quantities is inferred by the virtual measurement's algorithm from the database $\mathcal{D}$ and the common form \acp{pdf} are not assigned by the operator.

After the definition of the measurand, the input quantities and their probability distributions, finally the measurement model is designed.
The physical measurement model is constructed by the operator based on physical relationships or the execution of experiments or both.
For the virtual measurement, a suitable algorithm and its hyperparameter are selected.
The virtual measurement algorithm is then conditioned on the database $\mathcal{D}$ in the propagation stage.

\subsection*{Stage 2: Propagation}

In the context of \ac{gum}, the propagation stage obtains the \ac{pdf} of $Y$ by propagating the \acp{pdf} of the input quantities $X_i$ through the measurement model $f$.
For physical measurement, the \ac{gum} offers multiple propagation methods, which can result in different uncertainty estimates.
The choice of the propagation method depends on the measurement model and the input quantities.
The operator selects a suitable propagation method and executes the propagation.
The definition of the measurement model $f$ and the propagation of uncertainty are separated\footnote{For the physical measurement function $f$ \ac{ml} models can be used, \cf \cite[Sec.\~3.5]{muellerModellingComplexMeasurement2020}. However, model training and uncertainty propagation remain separate.}.
For virtual measurement, there are many algorithms available to implement the probabilistic measurement model $f$.
A popular choice is the \ac{vi} method, as presented in \cref{sec:virtual-measurement-uncertainty}.
The choice of algorithm depends on the properties of the data set $\mathcal{D}$.
The propagation of uncertainty is implicitly executed by the algorithm and the definition of the measurement model $f$, and the propagation of uncertainty occurs simultaneously during the conditioning of the model to the dataset (model training).
The operator may have to iterate different combinations of hyperparameters for the algorithm to obtain a well-tuned model.
Similar to the uncertainty of physical measurements, different algorithms can result in different uncertainty estimates.

For both measurement types, one obtains (an approximation of) the \ac{cdf} of $Y$.
The measurement uncertainty $U(Y)$ is then derived from the variance $\mathbb{V}[Y] = \sigma_Y^2$.
For the physical measurement, the uncertainty is an affixed value, valid for all future measurements, anticipating that all underlying assumptions hold.
For the virtual measurement, the uncertainty is a function of the realizations of the input quantities and changes for every new measurement.
If the underlying assumptions do not hold, a well-tuned algorithm will provide a large uncertainty estimate, indicating that the model is not reliable for the current measurement.
These differences must be taken into account in the documentation of the measurement uncertainty.

\subsection*{Stage 3: Documentation}

The documentation for the physical and the virtual measurements are equal in the sense that all information to reproduce the determination of uncertainty have to be reported.
For physical measurement, this includes the measurement model $f$, the input quantities $\bm{X}$, the \acp{pdf} of the input quantities, and the propagation method.
For the virtual measurement, this includes the algorithm, the hyperparameters, and the database $\mathcal{D}$. 
In both cases, the \ac{pdf} of the measurand $Y$ is documented.

The main difference between the two types of measurement is that the physical measurement uncertainty is a fixed value valid for all future measurements, whereas the virtual measurement's uncertainty is a function of the realizations of the input quantities individual to each measurement.
Therefore, for physical measurements, the \ac{pdf} is commonly approximated by a normal distribution $Y \sim \mathcal{N}\left(\hat{y}, \hat{\sigma}^2\right)$ and the parameters $\hat{y}$ and $\hat{\sigma}^2$ are reported.
For the virtual measurement, the parameters of the aforementioned normal distribution depend on the realizations of the input quantities and are therefore not fixed.
Instead, the conditioned model is stored and queried for realizations of the input quantities $\bm{x} \sim \bm{X}$, \ie $Y(\bm{x}) \sim \mathcal{N}\left(\hat{y}(\bm{x}), \hat{\sigma}^2(\bm{x})\right)$.
Coverage intervals are commonly reported based on the standard deviation $\sigma$ and the extension factor $k$.

\subsection*{Summary}

We discussed the association of virtual and physical measurement uncertainty and compared the stages of their determination analogous to the \ac{gum}.

In the formulation stage, both types of measurement agree on the definition of a measurand.
However, they differ in the selection of input quantities and the assignment of \acp{pdf} to these quantities.
The measurement model for physical measurements is constructed by the operator, whereas for virtual measurements, an algorithm and its hyperparameters are selected.

In the propagation stage, the \ac{pdf} of the measurand is obtained by propagating the \acp{pdf} of the input quantities through the measurement model.
For physical measurements, the \ac{gum} offers multiple propagation methods.
For virtual measurements, the algorithm implicitly executes the propagation of uncertainty during the conditioning of the model to the data set.
The uncertainty identified may vary depending on the choice of the propagation method or algorithm.

In the documentation stage, both types of measurement require all information necessary to reproduce the determination of uncertainty.
The main difference is that the physical measurement's uncertainty is a fixed value valid for all future measurements, whereas the virtual measurement's uncertainty is a function of the realizations of the input quantities individual to each measurement.

We conclude that the virtual measurement uncertainty determined by the procedure in \cref{sec:virtual-measurement-uncertainty} is equivalent to the physical measurement uncertainty as determined by the procedure in \cref{sec:physical-measurement-uncertainty}.
Therefore, the virtual measurement uncertainty can be used in the same way as the physical measurement uncertainty in quality management frameworks.
\section{Concluding Remarks}
\label{sec:concluding-remarks}

The virtual measurement is becoming more popular in industrial applications.
However, a proper definition of a virtual measurement and its associated uncertainty is lacking in the literature.
The virtual measurement uncertainty introduces a risk to all subsequent decisions and its use in established quality management frameworks remains limited.

We provided definitions of the virtual measured quantity value (\cf\cref{def:virt-measured-quant-value}), the virtual measurement uncertainty (\cf\cref{def:virt-measurement-uncertainty}), and the virtual measurement (\cf\cref{def:virt-measurement}) in line with the ratified \ac{gum}.
Furthermore, we outlined the three stages of uncertainty propagation for the physical measurement, as well as the virtual measurement following \cite[Sec.~5.1]{JCGM101}, before assessing the major similarities in \cref{sec:associate-uncertainty}.

Our work helps early adopters integrate virtual measurement in their established quality management workflow and be accurately informed regarding the risk of their downstream decisions.
Additionally, establishing virtual measurement is a first step towards standardization in the metrology community and broader acceptance in industry.

Our approach is limited to virtual measurements based on probabilistic models.
Yet, for deterministic (measurement) models the propagation of uncertainty carries two major difficulties: \begin{inparaenum}[$(i)$]
    \item the assignment of probability distributions to the input quantities, and
    \item the computational effort for propagation.
    \end{inparaenum}
Therefore, we promote the use of probabilistic models for virtual measurements.

The different methods for uncertainty propagation in \cite[Sec.~5.1]{JCGM101} have established boundary conditions and known limitations of their capabilities to propagate uncertainties.
For algorithms regarding probabilistic models, these boundary conditions are not established for virtual measurements.
There is evidence that different algorithms represent the uncertainty differently \eg \cite{schmahlingFrameworkBenchmarkingUncertainty2023}.
In the future, we will investigate whether conformalizing the probabilistic model as in \cite[Sec.~2.3]{angelopoulosGentleIntroductionConformal2022} will harmonize the results of different algorithms.
Another interesting research direction is the fusion of uncertainties in virtual measurements with the work of \citeauthor*{schmittMetrologicallyInterpretableFeature2022} on interpreting the latent vector of generative deep learning methods as metrological quantities.

\paragraph{Acknowledgments}
This research is funded by the Deutsche Forschungsgemeinschaft (DFG, German Research Foundation) under Germany’s Excellence Strategy — EXC-2023 Internet of Production — 390621612.

\printbibliography
\newpage
\appendix
\section{Appendix}
\label{sec:appendix}

\begin{definition}[Product]\label{def:product}
    A product is a description of the desired output of a production process \cite[chapter~3.7.6]{DINISO9000}.
\end{definition}

\begin{definition}[Produced Part]\label{def:produced_part}
    A produced part is an individual output of a production process. Therefore, a produced part is a realization or instance of a product.
\end{definition}

\begin{definition}[Database $\mathcal{D}$]\label{def:database}
    For every produced part $i \in \mathbb{N}^+$ we record process data $x_i$ and quality characteristics $y_i$ in a database $\mathcal{D} = \left\{ (x_i, y_i) | 0 < i < k  \right\}$ with $k+1$ entries.
\end{definition}

The definitions of $x_i$ and $y_i$ are analog to \cite[Def. 3-5]{cramerUncertaintyQuantificationBased2022}.

\begin{definition}
\label{def:process-data-bayes}
The \emph{process data} $\bm{X}$ is generated by $m \in \mathbb{N}$ sensors, where the sensor readings $s_j\, (0 \leq j<m)$ are given by a stochastic process (\cf \cref{def:stochasticprocess}).
Accordingly, the process data is modeled by $\bm{X}: T \times \Omega^m \xrightarrow[]{} S^m$ with $\bm{X}(t,\bar{\omega}):=[ s_j(t,\omega_j) ]^T$ where $\bar{\omega}:= [\omega_j]^T$.
\end{definition}

\begin{definition}
\label{def:quality-characterisitic-bayes}
The measurements of the \emph{quality characteristics} $Y: \Omega^n \xrightarrow[]{} \mathbb{R}^n$ are given by $n \in \mathbb{N}$ measurements, where every measurement $v_l\, (0\leq l<n)$ is a random variable $Y(\bar{\omega}): = [v_l(\omega_l)]^T$ where $\bar{\omega}:= [\omega_l]^T$.
\end{definition}

\begin{definition}
\label{def:stochasticprocess}
Let $u(t,\omega): T \times \Omega \xrightarrow[]{} S$ be a \emph{stochastic process}, where $t \in T \subset \mathbb{R}^+$ and $\omega \in \Omega$.
Here $\Omega$ is the sample space of the probability space $(\Omega, \mathcal{F}, P)$ with $\mathcal{F}$ being a $\sigma$-algebra and $P$ a probability measure.
\end{definition}

\subsection{Expectation Value}
\label{sec:expectation_value}
The expectation value of the posterior predictive density can be computed as follows
\begin{align}
    \mathbb{E}_{p(Y|\bm{X}, \mathcal{D}_{\bm{x}}, \mathcal{D}_{y})}[Y] &= \int_{\mathbb{R}} p(y|\bm{X}, \mathcal{D}_{\bm{x}}, \mathcal{D}_{y}) y \mathrm{d}y \\
    &= \int_{\mathbb{R}} \left(\int_{\mathbb{R}^P} p(\bm{w}|\mathcal{D}_{\bm{x}}, \mathcal{D}_{y}) p(y|\bm{X}, \bm{w}) \mathrm{d}\bm{w}\right) y \mathrm{d}y \label{eq:ex_val_def_post} \\ 
    &= \int_{\mathbb{R}^P} \int_{\mathbb{R}} p(\bm{w}|\mathcal{D}_{\bm{x}}, \mathcal{D}_{y}) p(y|\bm{X}, \bm{w})  y \mathrm{d}y \mathrm{d}\bm{w} \label{eq:ex_val_def_switch} \\ 
    &= \int_{\mathbb{R}^P} p(\bm{w}|\mathcal{D}_{\bm{x}}, \mathcal{D}_{y}) \left( \int_{\mathbb{R}} p(y|\bm{X}, \bm{w}) y \mathrm{d}y \right) \mathrm{d}\bm{w} \label{eq:ex_val_def_rearrange} \\ 
    &= \int_{\mathbb{R}^P} p(\bm{w}|\mathcal{D}_{\bm{x}}, \mathcal{D}_{y}) \mathbb{E}_{p(y|\bm{X}, \bm{w})}[y] \mathrm{d}\bm{w} \\
    &= \int_{\mathbb{R}^P} p(\bm{w}|\mathcal{D}_{\bm{x}}, \mathcal{D}_{y}) f(\bm{X}; \bm{w}) \mathrm{d}\bm{w} \label{eq:ex_val_def_reinsert} \\
    &= \mathbb{E}_{p(\bm{w}|\mathcal{D}_{\bm{x}}, \mathcal{D}_{y})}[f(\bm{X}; \bm{w})]. \label{eq:ex_val_def_final}
\end{align}
In \cref{eq:ex_val_def_post}, we inserted the definition of the posterior predictive density $p(Y|\bm{X}, \mathcal{D}_{\bm{x}}, \mathcal{D}_{y})$ from \cref{eq:posterior_predictive_distribution}. In \cref{eq:ex_val_def_switch}, we switched the order of integration. In \cref{eq:ex_val_def_reinsert}, we inserted the expectation value of the model density $p(Y|\bm{X}, \bm{w})$ from \cref{eq:exp_val_model_density} \cite[chapter~3.3]{Gal2016UncertaintyID}. 

\subsection{Variance}
\label{sec:variance}

For the derivation of the variance of the posterior predictive density, we first show a general property of the variance. The variance of a random variable $A \in  \mathbb{R}$ w.r.t. a probability density $p(A)$ can be expanded as follows
\begin{align}
    \mathbb{V}_{p(A)}[A] &= \mathbb{E}_{p(A)}[(A - \mathbb{E}_{p(A)}[A])^2] \\
    &= \mathbb{E}_{p(A)}[A^2 - 2A\mathbb{E}_{p(A)}[A] + (\mathbb{E}_{p(A)}[A])^2] \\
    &= \mathbb{E}_{p(A)}[A^2] - 2(\mathbb{E}_{p(A)}[A])^2 + (\mathbb{E}_{p(A)}[A])^2 \label{eq:identity_linearity} \\ 
    &= \mathbb{E}_{p(A)}[A^2] - (\mathbb{E}_{p(A)}[A])^2. \label{eq:identity_expansion}
\end{align}
In \cref{eq:identity_linearity}, we used the linearity property of the expectation value.

The variance of the posterior predictive density can be decomposed into two components by applying the expansion from \cref{eq:identity_expansion}
\begin{align}
    \mathbb{V}_{p(Y|\bm{X}, \mathcal{D}_{\bm{x}}, \mathcal{D}_{y})}[Y] &= \mathbb{E}_{p(Y|\bm{X}, \mathcal{D}_{\bm{x}}, \mathcal{D}_{y})}[Y^2] - (\mathbb{E}_{p(Y|\bm{X}, \mathcal{D}_{\bm{x}}, \mathcal{D}_{y})}[Y])^2 \label{eq:var_def_definition} \\ 
    &= \mathbb{E}_{p(Y|\bm{X}, \mathcal{D}_{\bm{x}}, \mathcal{D}_{y})}[Y^2] - (\mathbb{E}_{p(\bm{w}|\mathcal{D}_{\bm{x}}, \mathcal{D}_{y})}[f(\bm{X}; \bm{w})])^2 \label{eq:var_def_insert}.
\end{align}
In \cref{eq:var_def_insert}, we inserted the expectation value of the posterior predictive density from \cref{eq:ex_val_def_final}.

The first component of \cref{eq:var_def_insert} can be computed as follows
\begin{align}
    \mathbb{E}_{p(Y|\bm{X}, \mathcal{D}_{\bm{x}}, \mathcal{D}_{y})}[Y^2] &= \int_{\mathbb{R}} p(y|\bm{X}, \mathcal{D}_{\bm{x}}, \mathcal{D}_{y}) Y^2 \mathrm{d}y \\
    &= \int_{\mathbb{R}} \left(\int_{\mathbb{R}^P} p(\bm{w}|\mathcal{D}_{\bm{x}}, \mathcal{D}_{y}) p(y|\bm{X}, \bm{w}) \mathrm{d}\bm{w}\right) y^2 \mathrm{d}y \label{eq:ex1_insert} \\
    &= \int_{\mathbb{R}^P} \int_{\mathbb{R}} p(\bm{w}|\mathcal{D}_{\bm{x}}, \mathcal{D}_{y}) p(y|\bm{X}, \bm{w}) y^2 \mathrm{d}y \mathrm{d}\bm{w} \label{eq:ex1_switch} \\
    &= \int_{\mathbb{R}^P} p(\bm{w}|\mathcal{D}_{\bm{x}}, \mathcal{D}_{y}) \left(\int_{\mathbb{R}}  p(y|\bm{X}, \bm{w}) y^2 \mathrm{d}y \right) \mathrm{d}\bm{w} \label{eq:ex1_rearrange} \\
    &= \int_{\mathbb{R}^P} p(\bm{w}|\mathcal{D}_{\bm{x}}, \mathcal{D}_{y}) \mathbb{E}_{p(y|\bm{X}, \bm{w})}[y^2] \mathrm{d}\bm{w} \\
    &= \int_{\mathbb{R}^P} p(\bm{w}|\mathcal{D}_{\bm{x}}, \mathcal{D}_{y}) \left( \mathbb{E}_{p(y|\bm{X}, \bm{w})}[(y - \mathbb{E}_{p(y|\bm{x}, \bm{w})}[y])^2] + (\mathbb{E}_{p(y|\bm{X}, \bm{w})}[y])^2 \right) \mathrm{d}\bm{w} \label{eq:ex1_identity} \\
    &= \int_{\mathbb{R}^P} p(\bm{w}|\mathcal{D}_{\bm{x}}, \mathcal{D}_{y}) \left( \sigma_n^2(\bm{X}; \bm{w}) + f^2(\bm{X}; \bm{w}) \right) \mathrm{d}\bm{w} \label{eq:ex1_reinsert} \\
    &= \int_{\mathbb{R}^P} p(\bm{w}|\mathcal{D}_{\bm{x}}, \mathcal{D}_{y}) \sigma_n^2(\bm{X}; \bm{w}) \mathrm{d}\bm{w} + \int_{\mathbb{R}^P} p(\bm{w}|\mathcal{D}_{\bm{x}}, \mathcal{D}_{y}) f^2(\bm{X}; \bm{w})  \mathrm{d}\bm{w} \\
    &= \mathbb{E}_{p(\bm{w}|\mathcal{D}_{\bm{x}}, \mathcal{D}_{y})}[\sigma_n^2(\bm{X}; \bm{w})] + \mathbb{E}_{p(\bm{w}|\mathcal{D}_{\bm{x}}, \mathcal{D}_{y})}[f^2(\bm{X}; \bm{w})]. \label{eq:ex1_final}
\end{align}
In \cref{eq:ex1_insert}, we inserted the definition of the posterior predictive density $p(Y|\bm{X}, \mathcal{D}_{\bm{x}}, \mathcal{D}_{y})$ from \cref{eq:posterior_predictive_distribution}. In \cref{eq:ex1_switch}, we switched the order of integration. In \cref{eq:ex1_identity}, we applied the expansion of the variance from \cref{eq:identity_expansion}. In \cref{eq:ex1_reinsert}, we inserted the expectation value and variance of the model density $p(Y|\bm{X}, \bm{w})$ from \cref{eq:exp_val_model_density} and \cref{eq:var_model_density}.

Inserting our result for the component part from \cref{eq:ex1_final} in \cref{eq:var_def_insert}, the variance of the posterior predictive density is given by
\begin{align}
    \mathbb{V}_{p(Y|\bm{X}, \mathcal{D}_{\bm{x}}, \mathcal{D}_{y})}&[Y] \\
    = \mathbb{E}&_{p(\bm{w}|\mathcal{D}_{\bm{x}}, \mathcal{D}_{y})}[\sigma_n^2(\bm{X}; \bm{w})] + \mathbb{E}_{p(\bm{w}|\mathcal{D}_{\bm{x}}, \mathcal{D}_{y})}[f^2(\bm{X}; \bm{w})] - (\mathbb{E}_{p(\bm{w}|\mathcal{D}_{\bm{x}}, \mathcal{D}_{y})}[f(\bm{X}; \bm{w})])^2 \\
    = \mathbb{E}&_{p(\bm{w}|\mathcal{D}_{\bm{x}}, \mathcal{D}_{y})}[\sigma_n^2(\bm{X}; \bm{w})] + \mathbb{V}_{p(\bm{w}|\mathcal{D}_{\bm{x}}, \mathcal{D}_{y})}[f(\bm{X}; \bm{w})]. \label{eq:var_def_inverse_reinsert}
\end{align}
In \cref{eq:var_def_inverse_reinsert}, we applied the expansion of the variance from \cref{eq:identity_expansion} \cite[chapter~3]{kwonUncertaintyQuantificationUsing}.

\subsection{Kullback-Leibler Divergence}
\label{sec:kl_divergence}

A fundamental characteristic of Bayesian inference is that the model parameters are treated probabilistically. Initially, we assume a prior density $p(\bm{w})$ for the model parameters $\bm{w}$. Given a data set $\mathcal{D}$ with $D \in \mathbb{N}$ independent observations of the production process
\begin{align}
    \mathcal{D} = \{(\bm{x}^{(1)}, y^{(1)}), \dots, (\bm{x}^{(D)}, y^{(D)})\},
\end{align}
the prior density $p(\bm{w})$ is updated to the posterior density $p(\bm{w}|\mathcal{D}_{\bm{x}}, \mathcal{D}_{y})$, where $\mathcal{D}_{\bm{x}} = \{\bm{x}^{(1)}, \dots, \bm{x}^{(D)}\}$ and $\mathcal{D}_{y} = \{y^{(1)}, \dots, y^{(D)}\}$. The update to the posterior density is specified by Bayes' theorem
\begin{align}
    p(\bm{w}|\mathcal{D}_{\bm{x}}, \mathcal{D}_{y}) = \frac{p(\mathcal{D}_{y}|\mathcal{D}_{\bm{x}}, \bm{w})p(\bm{w})}{p(\mathcal{D}_{y}|\mathcal{D}_{\bm{x}})}, \label{eq:bayes_theorem}
\end{align}
where we defined the likelihood function $p(\mathcal{D}_{y}|\mathcal{D}_{\bm{x}}, \bm{w})$ by
\begin{align}
    p(\mathcal{D}_{y}|\mathcal{D}_{\bm{x}}, \bm{w}) = \prod_{d=1}^D p(y^{(d)}|\bm{x}^{(d)}, \bm{w}). \label{eq:likelihood_function}
\end{align}
The likelihood function factorizes over the single observations. The reason is that we assumed statistical independence of the observations in the data set $\mathcal{D}$. Finally, the evidence density $p(\mathcal{D}_{y}|\mathcal{D}_{\bm{x}})$ provides the proper normalization of the posterior density in Bayes' theorem \cite[Chap.~2]{bishopPatternRecognitionMachine2006}.

The \ac{kl} divergence between variational and posterior density can be rewritten as follows
\begin{align}
    \mathrm{KL}&[q(\bm{w}; \bm{\phi})||p(\bm{w}|\mathcal{D}_{\bm{x}}, \mathcal{D}_{y})] \\ &= \mathbb{E}_{q(\bm{w}; \bm{\phi})}\left[ \log \frac{q(\bm{w}; \bm{\phi})}{p(\bm{w}|\mathcal{D}_{\bm{x}}, \mathcal{D}_{y})} \right] \label{eq:kl_definition} \\
    &= \int_{\mathbb{R}^P} q(\bm{w}; \bm{\phi}) \log \frac{q(\bm{w}; \bm{\phi})}{p(\bm{w}|\mathcal{D}_{\bm{x}}, \mathcal{D}_{y})} \mathrm{d}\bm{w} \\
    &= \int_{\mathbb{R}^P} q(\bm{w}; \bm{\phi}) \log \frac{q(\bm{w}; \bm{\phi})p(\mathcal{D}_{y}|\mathcal{D}_{\bm{x}})}{p(\bm{w})p(\mathcal{D}_{\bm{x}}, \mathcal{D}_{y}|\bm{w})} \mathrm{d}\bm{w} \label{eq:kl_bayes_theorem} \\
    &= \int_{\mathbb{R}^P} q(\bm{w}; \bm{\phi}) \left( \log \frac{q(\bm{w}; \bm{\phi})}{p(\bm{w})} + \log p(\mathcal{D}_{y}|\mathcal{D}_{\bm{x}}) - \log p(\mathcal{D}_{\bm{x}}, \mathcal{D}_{y}|\bm{w}) \right) \mathrm{d}\bm{w} \\
    &\begin{aligned}
        &= \int_{\mathbb{R}^P} q(\bm{w}; \bm{\phi}) \log \frac{q(\bm{w}; \bm{\phi})}{p(\bm{w})} \mathrm{d}\bm{w} \\
        &\hskip8em\relax + \int_{\mathbb{R}^P} q(\bm{w}; \bm{\phi}) \log p(\mathcal{D}_{\bm{x}}, \mathcal{D}_{y})\mathrm{d}\bm{w} \\
        &\hskip16em\relax - \int_{\mathbb{R}^P} q(\bm{w}; \bm{\phi}) \log p(\mathcal{D}_{\bm{x}}, \mathcal{D}_{y}|\bm{w}) \mathrm{d}\bm{w}
    \end{aligned}\\
    &= \mathbb{E}_{q(\bm{w}; \bm{\phi})}\left[\log \frac{q(\bm{w}; \bm{\phi})}{p(\bm{w})}\right] + \log p(\mathcal{D}_{\bm{x}}, \mathcal{D}_{y}) - \mathbb{E}_{q(\bm{w}; \bm{\phi})}[ \log p(\mathcal{D}_{\bm{x}}, \mathcal{D}_{y}|\bm{w}) ] \label{eq:kl_normalization_surrogate} \\
    &= \mathrm{KL}[q(\bm{w}; \bm{\phi})||p(\bm{w})] + \log p(\mathcal{D}_{\bm{x}}, \mathcal{D}_{y}) - \mathbb{E}_{q(\bm{w}; \bm{\phi})}[ \log p(\mathcal{D}_{\bm{x}}, \mathcal{D}_{y}|\bm{w}) ] \label{eq:kl_result}
\end{align}
In \cref{eq:kl_definition}, we used the definition of the \ac{kl} divergence from \cref{eq:def_kl}. In \cref{eq:kl_bayes_theorem}, we applied Bayes' theorem from \cref{eq:bayes_theorem}. In \cref{eq:kl_normalization_surrogate}, we used the normalization of the variational density $q(\bm{w}; \bm{\phi})$ to $1$ to compute the integral in the second term.

Therefore, the \ac{kl} divergence from \cref{eq:kl_result} is minimized by the optimal variational variables
\begin{align}
    \bm{\phi}^* &= \argmin_{\bm{\phi} \in \Phi} \mathrm{KL}[q(\bm{w}; \bm{\phi})||p(\bm{w}|\mathcal{D}_{\bm{x}}, \mathcal{D}_{y})] \\
    &= \argmin_{\bm{\phi} \in \Phi} \left( \mathrm{KL}[q(\bm{w}; \bm{\phi})||p(\bm{w})] -  \mathbb{E}_{q(\bm{w}; \bm{\phi})}[ \log p(\mathcal{D}_{\bm{x}}, \mathcal{D}_{y}|\bm{w}) ]\right), \label{eq:kl_independence}
\end{align}
where the second term of \cref{eq:kl_result} vanishes due to its independence of the variational variables $\bm{\phi}$ \cite[chapter~3]{blundellWeightUncertaintyNeural2015}.
\end{document}